\documentclass[twocolumn,prl,reprint,twocolumn,superscriptaddress,amsmath,amssymb]{revtex4-1}
\usepackage[latin9]{inputenc}
\usepackage{array}
\usepackage{float}
\usepackage{multirow}
\usepackage{amsmath}
\usepackage{graphicx}
\usepackage[unicode=true,pdfusetitle,
 bookmarks=false,
 breaklinks=false,pdfborder={0 0 1},backref=false,colorlinks=false]
 {hyperref}

\makeatletter

\providecommand{\tabularnewline}{\\}

\usepackage{color}
\usepackage[caption=false]{subfig}

\usepackage{graphicx}

\makeatother

\begin{document}
\title{Fractional Conductance in Strongly Interacting 1D Systems}
\author{Gal Shavit}
\affiliation{Department of Condensed Matter Physics, Weizmann Institute of Science,
Rehovot, Israel 76100}
\author{Yuval Oreg}
\affiliation{Department of Condensed Matter Physics, Weizmann Institute of Science,
Rehovot, Israel 76100}
\begin{abstract}
We study one dimensional clean systems with few channels and strong
electron-electron interactions. We find that in several circumstances,
even when time reversal symmetry holds, they may lead to two terminal
fractional quantized conductance and fractional shot noise. The condition
on the commensurability of the Fermi momenta of the different channels
and the strength of interactions resulting in such remarkable phenomena
are explored using abelian bosonization. Finite temperature and length
effects are accounted for by a generalization of the Luther-Emery
re-fermionization at specific values of the interaction strength.
We discuss the connection of our model to recent experiments in confined
2DEG, featuring possible fractional conductance plateaus. One of the
most dominant observed fractions, with two terminal conductance equals
to $\frac{2}{5}\frac{e^{2}}{h}$, is found in several scenarios of
our model. Finally, we discuss how at very small energy scales the
conductance returns to an integer value and the role of disorder.
\end{abstract}
\maketitle
\emph{Introduction and main results.---} Fractional quantum Hall
(FQH) effect, exhibiting a fractionally quantized value of the Hall
conductance in units of $\frac{e^{2}}{h}$ \cite{IntroFQHexp1,IntroFQHexp2},
is a hallmark of strongly correlated electron systems, featuring composite
particles, fractionally charged excitations, and fractional exchange
statistics \cite{IntroFQLaughlin,IntroFQHaldane,IntroFQHJain}. In
recent years, theoretical studies of very clean one-dimensional (1D)
quantum nano-wires with broken time reversal symmetry predict fractional
values of the two-terminal conductance \cite{HelicalSelaOreg}, as
well as fractional shot noise \cite{SelaShotNoise}. In contrast to
the quantum Hall effect the one-dimensional wires are not topologically
protected from the effect of impurity scattering, and hence observation
of approximate fractional conductance and shot noise requires high
degrees of purity.

Interest in such fractional states has risen recently, with experimental
evidence for fractional transport in split-gate 1D constrictions made
in germanium two-dimensional layers \cite{SelfOrgaanized}, and in
GaAs/AlGaAs heterostructures, even in the absence of an external magnetic
field \cite{PepperArxiv}. Strong interactions between the quasi-particles
in 1D are expected to play an important role in determining transport
properties, especially when the electronic confinement in the transverse
direction is somewhat relaxed \cite{WignerReview,Wigner2}.

In this manuscript, we explore a two-band \footnote{We generalized the two band model to the multi-mode case in the SM
Sec. \ref{NbandsGeneralization}} fermionic 1D system, that bands for example could be, but not necessarily
are, the spin degree of freedom. We find that even in the absence
of time-reversal breaking the combination of tuning of the chemical
potentials of the bands, and very strong inter-band interactions,
leads to universal fractional transport properties at intermediate,
and experimentally relevant, energy scales. We argue how in the very
clean case at ultra-small temperatures, the conductance recovers an
integer value. The role of disorder is discussed in the supplementary
materials (SM) Sec. \ref{ImpuritiesSec}.

We perform finite temperature and length analysis of the two-terminal
conductance, employing RG analysis procedure and re-fermionization
at specific values of the interaction that generalizes the Luther-Emery
point \cite{LutherEmery}. Finally, we use our novel results to suggest
plausible scenarios that fit reported measurements, including conductance
equals to $\frac{2}{5}$ in units of $\frac{e^{2}}{h}$; which is
one of the most experimentally predominant fractions \cite{PepperArxiv}.

\begin{figure}
\begin{centering}
\includegraphics[scale=0.3]{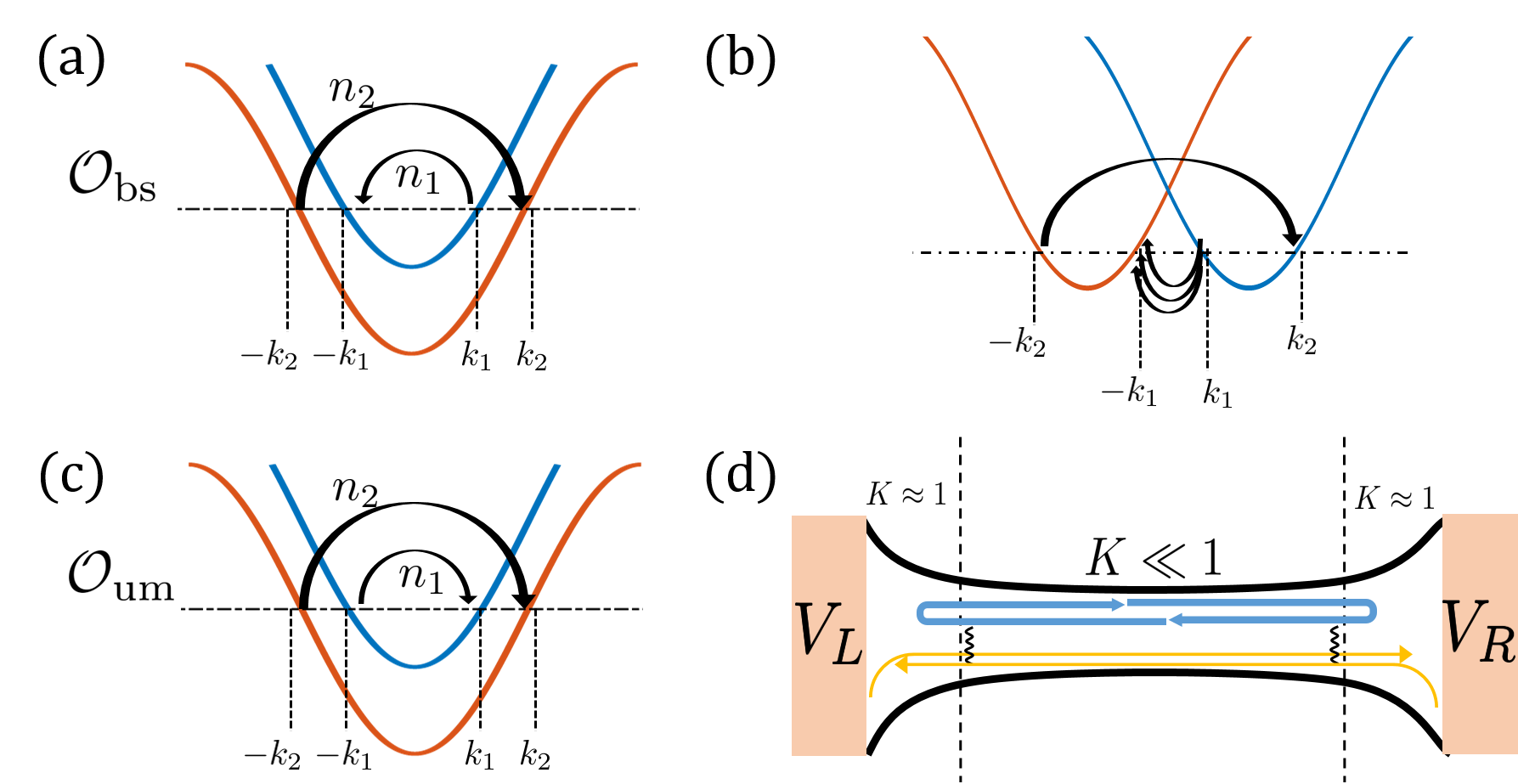}
\par\end{centering}
\caption{\label{fig:Fig1Schem} (a) Two-band dispersion with an example of
a backscatering $\mathcal{O}_{{\rm bs}}$ process which conserves
momentum when the chemical potential (horizontal dashed line) is such
that $n_{1}k_{1}=n_{2}k_{2}$. (b) An example of a time-reversal invariant
backscattering process, $\left(n_{1},n_{2}\right)=\left(3,1\right)$,
occurring for fractional filling of a Rashba nano-wire, see SM Sec.
\ref{TrSec} \cite{Supplement}. (c) Similarly to (a), an umklapp
process $\mathcal{O}_{\mathrm{um}}$ with a net momentum change, conserves
lattice momentum when $n_{1}k_{1}+n_{2}k_{2}=\pi\cdot\mathrm{integer}$,
stabilizing a fractional Mott-insulator phase. (d) Illustration of
a scenario where both bands interact throughout out the wire, yet
one is confined and does not reach the reservoirs. This leads to a
variety of possible fractional conductance values, see SM Sec. \ref{NbandsGeneralization}
\cite{Supplement}.}
\end{figure}

At the core of our analysis is the observation that when the electro-chemical
potential $\mu$ is tuned properly, the backscattering momentum of
$n_{1}$ right moving electrons at the Fermi level, is compensated
by backward scattering of $n_{2}$ left moving electrons (and vice
versa), so that multi-electron scattering processes occur in a clean
momentum conserving system (see Fig. \ref{fig:Fig1Schem}a). Such
processes are relevant in the RG sense when interactions inside the
wire are sufficiently strong. Remarkably, time-reversal symmetry is
not necessarily broken when $\left|n_{1}+n_{2}\right|$ is even (see
SM Sec. \ref{TrSec}). In the presence of a lattice umklapp processes
may occur, they are formally accounted for by changing the relative
sign of $n_{1},n_{2}$, see Fig. \ref{fig:Fig1Schem}c.

\emph{Theoretical model.---} We consider a 1D system which hosts
two interacting electron species, with annihilation operators $c_{1}\left(x\right)$
and $c_{2}\left(x\right)$ at position $x$ and different chemical
potentials $\mu_{i}$. The Hamiltonian is
\begin{align}
H & =\int dx\,c_{i}^{\dagger}\left(x\right)\left[\delta_{ij}\left(-\mu_{i}-\frac{\partial_{x}^{2}}{2m_{i}}\right)\right]c_{j}\left(x\right)\nonumber \\
 & +\int dx\int dx'\rho_{i}\left(x\right)U_{ij}\left(x,x'\right)\rho_{j}\left(x'\right),\label{eq:ModelHfermions}
\end{align}
where $\rho_{i}=c_{i}^{\dagger}c_{i}$, $U_{ij}$ is the interaction
matrix, and summation over repeated indices, $i,j=1,2$, is implied.
The model \eqref{eq:ModelHfermions} is conveniently analyzed in the
framework of abelian bosonization \cite{giamarchi2004quantum,1dFL}.
Linearizing the spectrum around the Fermi energy, the fermionic operators
are decomposed into chiral modes, such that $c_{i}=\psi_{R}^{i}+\psi_{L}^{i}$,
with $R$ ($L$) being the right (left) moving mode. These are then
represented in terms of bosonic variables 
\begin{equation}
\psi_{r}^{i}\sim\frac{1}{\sqrt{2\pi a}}e^{irk_{i}x}e^{-i\left(r\phi_{i}-\theta_{i}\right)},\label{eq:bosonization}
\end{equation}
with $k_{i}$ the Fermi momentum of species $i$, $a$ is a short-distance
cutoff, $r=+1$ ($-1$) for right (left) movers, and the bosonic variables
satisfy the algebra $\left[\phi_{i}\left(x\right),\partial_{x}\theta_{j}\left(x'\right)\right]=i\pi\delta_{ij}\delta\left(x-x'\right).$
The operator $-\frac{1}{\pi}\partial_{x}\phi_{i}$ ($\frac{1}{\pi}\partial_{t}\phi_{i}$)
represents the normally-ordered charge (current) density of the $i$
species. The forward scattering part of the interaction $U$ is incorporated
into the Hamiltonian $H_{{\rm fs}}$ by employing proper Luttinger
parameters, and diagonalized by defining $\phi_{\pm}=\frac{1}{\sqrt{2}}\left(\phi_{1}\pm\phi_{2}\right)$
and $\theta_{\pm}=\frac{1}{\sqrt{2}}\left(\theta_{1}\pm\theta_{2}\right)$,
such that 
\begin{equation}
H_{{\rm fs}}=\sum_{\eta=\pm}\frac{u_{\eta}}{2\pi}\int dx\left[\frac{1}{K_{\eta}(x)}\left(\partial_{x}\phi_{\eta}\right)^{2}+K_{\eta}(x)\left(\partial_{x}\theta_{\eta}\right)^{2}\right].\label{eq:LuttingerPM}
\end{equation}
Note that whereas the $\left(+\right)$ sector in \eqref{eq:LuttingerPM}
corresponds to the total charge sector, the $\left(-\right)$ does
not \textit{necessarily} represent spin. The distinction between different
species is kept general at this point. The Luttinger parameters may
be evaluated for weak interactions yielding: $g\equiv\frac{U_{0}}{\pi v_{F}}$,
$K_{\pm}\approx\sqrt{\frac{1+g\frac{1\mp1}{2}}{1+g\frac{1\pm3}{2}}}$,
and $u_{\pm}\approx\frac{v_{F}}{2}\sqrt{\left(2+g\right)\left(2+g\pm2g\right)}$
with $U_{q}$ the Fourier transform of the interaction, and $v_{F}$
the Fermi velocity\cite{giamarchi2004quantum}. We shall henceforth
assume for simplicity that $u_{+}\approx u_{-}\equiv u$, and that
the Luttinger liquid parameter is spatially smooth (on a scale of
$1/k_{i})$.

We now consider backscattering interactions which involve both species.
Generally, $\mu_{1}\neq\mu_{2}$, and we neglect processes that do
not conserve momentum. The operator $\mathcal{O}_{{\rm bs}}\sim\left(\psi_{R}^{1\dagger}\psi_{L}^{1}\right)^{\alpha}\left(\psi_{L}^{2}\psi_{R}^{\dagger2}\right)^{\beta}$
\cite{FermionicPower} is potentially relevant when $\alpha k_{1}\approx\beta k_{2}$
and nullified otherwise, due to the integral on coordinate $x$ (cf.
Fig. \ref{fig:Fig1Schem}a). Similarly, in the presence of external
periodic potential an umklapp type process$\mathcal{O}_{{\rm um}}\sim\left(\psi_{R}^{1\dagger}\psi_{L}^{1}\right)^{\alpha}\left(\psi_{R}^{2}\psi_{L}^{\dagger2}\right)^{\beta}$
may be relevant when $\alpha k_{1}+\beta k_{2}\approx\pi\cdot\mathrm{integer}$
and the lattice momentum is conserved (Fig. \ref{fig:Fig1Schem}c).
In Rashba nano-wires (cf. Fig. \ref{fig:Fig1Schem}b) or in case of
electron and holes bands, the right movers (and also the left movers)
of different species have opposite sign of Fermi momentum, then ${\cal O_{{\rm um}}}$conserves
momentum even in the absence of a lattice when $\alpha k_{1}\approx\beta k_{2}$
\cite{Rashbanano} (notice that in the Rashba nano-wires species are
identified by their helicity). We neglect several additional processes
that can be ruled out when two species are spatially separated, when
the Fermi momenta mismatch considerably, or due to strong repulsive
interactions which suppress (momentum conserving) pair hopping.

We may write a general scattering operator using the bosonized fields
\begin{equation}
\mathcal{O}_{\lambda}^{n_{1},n_{2}}=\int dx\frac{\lambda}{\left(2\pi\right)^{\left|n_{1}\right|+\left|n_{2}\right|}}\cos\left[2\left(n_{1}\phi_{1}+n_{2}\phi_{2}\right)\right],\label{eq:GenOnm}
\end{equation}
with the coupling strength $\lambda\propto\left(U_{2k_{1}}\right)^{\left|n_{1}\right|}\left(U_{2k_{2}}\right)^{\left|n_{2}\right|}$.
The integers $n_{i}$ have the opposite (same) sign for backscattering-
$\mathcal{O}_{{\rm bs}}$ (umklapp- $\mathcal{O}_{{\rm um}}$) processes.
The relevance of $\mathcal{O}_{\lambda}^{n_{1},n_{2}}$, in an RG
sense, can be understood by treating $\lambda$ as a small perturbation
compared to \eqref{eq:LuttingerPM}. At tree-level, the RG flow is
$\frac{d\lambda}{dl}=\left(2-D\right)\lambda$, with $l$ the flow
parameter, and the scaling dimension 
\begin{equation}
D=\left(n_{1}^{2}+n_{2}^{2}\right)\frac{K_{+}+K_{-}}{2}+n_{1}n_{2}\left(K_{+}-K_{-}\right).\label{eq:dimension}
\end{equation}
Therefore, the relevance condition $D<2$ can be met for \textit{sufficiently
strong repulsive interacti}ons. As $\mathcal{O}_{\lambda}^{n_{1},n_{2}}$
flows to strong coupling, a gap opens up in the sector $\phi_{g}\equiv\frac{n_{1}\phi_{1}+n_{2}\phi_{2}}{\sqrt{n_{1}^{2}+n_{2}^{2}}}$,
given by $\Delta_{\lambda}\approx ty^{\frac{1}{2-D}}$, with $t$
a typical bandwidth, and the dimensionless coupling strength $y\equiv\lambda\frac{\left(2\pi\right)^{1-\left|n_{1}\right|-\left|n_{2}\right|}}{u}$.
For temperatures above $T^{*}\equiv\Delta_{\lambda}$, or for lengths
shorter than $L^{*}\equiv\frac{u}{\Delta_{\lambda}}$, the RG flow
is cut-off before reaching strong coupling, and one finds the gap
$\Delta_{\lambda}$ scales as $\sim T^{D-1}$ or $\sim L^{1-D}$,
respectively.

\emph{Fractional two-terminal conductance.---} A setup in which the
1d system is smoothly connected (on the scale of $k_{1,2}^{-1}$)
at its ends to non-interacting reservoirs is considered. We begin
by considering a scattering problem, in the spirit of \cite{HelicalSelaOreg}.
By defining chiral bosonic fields $\varphi_{r}^{i}=\frac{\theta_{i}-r\phi_{i}}{\sqrt{2}}$,
we construct an incoming current vector $\vec{I}=\left(I_{R,1},I_{R,2},I_{L,1},I_{L,2}\right)^{T}$
with $I_{r,i}=\frac{e}{2\pi}\partial_{t}\varphi_{r}^{i}|_{x=r\infty}$,
and similarly an outgoing vector $\vec{O}$ with $O_{r,i}=\frac{e}{2\pi}\partial_{t}\varphi_{r}^{i}|_{x=-r\infty}$.
In the limit $\lambda\rightarrow\infty$, $\phi_{g}$ is gapped inside
the system, thus current flowing in this channel is fully backscattered,
i.e., $\sum_{i}n_{i}\partial_{t}\left(\varphi_{L}^{i}-\varphi_{R}^{i}\right)=0$.
In the sector orthogonal to $\phi_{g}$, $\phi_{f}\equiv\frac{n_{2}\phi_{1}-n_{1}\phi_{2}}{\sqrt{n_{1}^{2}+n_{2}^{2}}}$,
the current is unobstructed (in a clean wire), and we may write $\partial_{t}\left[n_{2}\varphi_{r}^{1}-n_{1}\varphi_{r}^{2}\right]_{x=\infty}=\partial_{t}\left[n_{2}\varphi_{r}^{1}-n_{1}\varphi_{r}^{2}\right]_{x=-\infty}$.
Using these conditions, we find the scattering matrix connecting the
current vectors $\vec{O}=S\vec{I}$ and the two-terminal conductance
$g\equiv\frac{h}{e^{2}}G$ (see SM Sec. \ref{NbandsGeneralization}
\cite{Supplement}), 
\begin{equation}
g=\frac{\left(n_{1}-n_{2}\right)^{2}}{n_{1}^{2}+n_{2}^{2}}.\label{eq:FracG}
\end{equation}
Thus, we find a myriad of possible fractionalized $g$ values. These
are \textit{universal}, in that they do not depend on details of the
model, e.g., the strength of interactions, and rely solely on $\lambda$
flowing to strong coupling limit, and on taking the limits $L\rightarrow\infty$,
$T\rightarrow0$. (Notice that by taking $n_{2}=n_{1}+1$, the fractional
values for the helical wire discussed in Refs. \cite{HelicalSelaOreg,LossHelical}
are obtained.)

One may consider additional 1D transport scenarios. A Coulomb drag
setup \cite{SternColumbDrag} in which the Fermi levels of the different
wires is commensurate in a similar manner will also lead to a \textit{fractional
transconductance} $g_{12}$. A situation when the species $i$ is
confined to the wire, i.e., does not couple to the leads, yet still
strongly interacts with species $\bar{i}$ (see Fig. \ref{fig:Fig1Schem}d),
would result in a different measured coefficient $g_{i{\rm c}}$.
Using the same scattering approach, one finds
\begin{equation}
g_{12}=-\frac{n_{1}n_{2}}{n_{1}^{2}+n_{2}^{2}},\,\,\,\,\,g_{i{\rm c}}=\frac{n_{i}^{2}}{n_{1}^{2}+n_{2}^{2}}.\label{eq:OtherGs}
\end{equation}

\emph{Generalized Luther-Emery line.---} We now wish to understand
the behavior of the fractional conductance in a finite temperature
and/or length. One expects the asymptotic value \eqref{eq:FracG}
to hold well-below $T^{*}$, whereas for sufficiently high $T$, the
gap renormalization will lead to power-law corrections to the integer
value $g=2$ (and similarly for $\frac{1}{L^{*}}$). We begin our
calculation by imposing boundary conditions at the connection of the
system to the leads, accounting for the interactions in the system
bulk \cite{LeadsBC},
\begin{equation}
\left[\frac{u_{\eta}}{K_{\eta}^{2}}\partial_{x}\pm\partial_{t}\right]\phi_{\eta}\left(x=\pm\frac{L}{2}\right)=\frac{1+\eta}{\sqrt{2}}\int dEf\left(E\pm\frac{V}{2}\right),\label{eq:BCradiative}
\end{equation}
with $\eta=\pm$, giving us a total of four equations. We use the
full Hamiltonian $H=H_{{\rm fs}}+\mathcal{O}_{\lambda}^{n_{1},n_{2}}$
to write our action in terms of $\phi_{g}$ and $\phi_{f}$ sectors
and their cross interactions, see SM Sec. \ref{RefermionizationDetails}
\cite{Supplement}. Upon shifting $\phi_{f}\rightarrow\phi_{f}+Q\phi_{g}$
(with $Q$ an appropriate constant), we neglect irrelevant cross terms,
and re-scale the bosonic fields $\phi_{g,f}\rightarrow\tilde{\phi}_{g,f}$
such that (i) the $\tilde{\phi}_{f}$ sector is non-interacting, and
(ii) the backscattering term is written in a form $\sim y\cos\left(2\tilde{\phi}_{g}\right)$.
We thus find that for given values of $n_{1,2}$, there exists a line
in the $K_{+}$-$K_{-}$ plane where the $\tilde{\phi}_{g}$ sector
is quadratic in fermionic variables, and the entire Hamiltonian may
be re-fermionized. This line is a novel generalization of the well-known
Luther-Emery point \cite{LutherEmery}.

Upon re-fermionization, Eq. \eqref{eq:BCradiative} may be solved
as a set of linear equations in the limit of adiabatically formed
gap \cite{AdiabaticGap}, and we find the total charge current $j_{c}=\frac{\sqrt{2}}{\pi}\partial_{t}\phi_{+}$
\cite{Supplement}
\begin{align}
j_{c}\left(E\right) & =2\frac{1}{2\pi}\delta f\Theta\left(E-\Delta\right)\nonumber \\
 & +\frac{1}{2\pi}\delta f\Theta\left(\Delta-E\right)\frac{\left(n_{1}-n_{2}\right)^{2}+2\chi\left(L\right)}{n_{1}^{2}+n_{2}^{2}+\chi\left(L\right)},\label{eq:jcGeneral}
\end{align}
with $\Delta=ty$, $\chi\left(L\right)\propto\sinh^{-2}\frac{\Delta}{u}L$
and $\delta f=f\left(E-\frac{V}{2}\right)-f\left(E+\frac{V}{2}\right)$.
Integrating over energy and restoring units, one obtains the result
\eqref{eq:FracG} in the limit $T\rightarrow0$, $L\rightarrow\infty$.
For temperatures well above the gap, \eqref{eq:jcGeneral} implies
a correction to the conductance $\delta g\equiv2-g$ with $\delta g\propto\Delta/T$.
Similarly for lengths much shorter than $L^{*}$, $\delta g\propto\left(\Delta L/u\right)^{2}$.
We may infer the power-laws for $\delta g$ slightly away from the
re-fermionization line (where the $\tilde{\phi}_{g}$ sector has weak
interactions) from the renormalization of the gap, to obtain
\begin{equation}
\delta g\propto\left(T/\text{\ensuremath{\Delta}}\right)^{D-2},\,\,\left(\Delta L/u\right)^{4-2D}.\label{eq:deltaGpowerLaw}
\end{equation}

\begin{figure}
\includegraphics[scale=0.45]{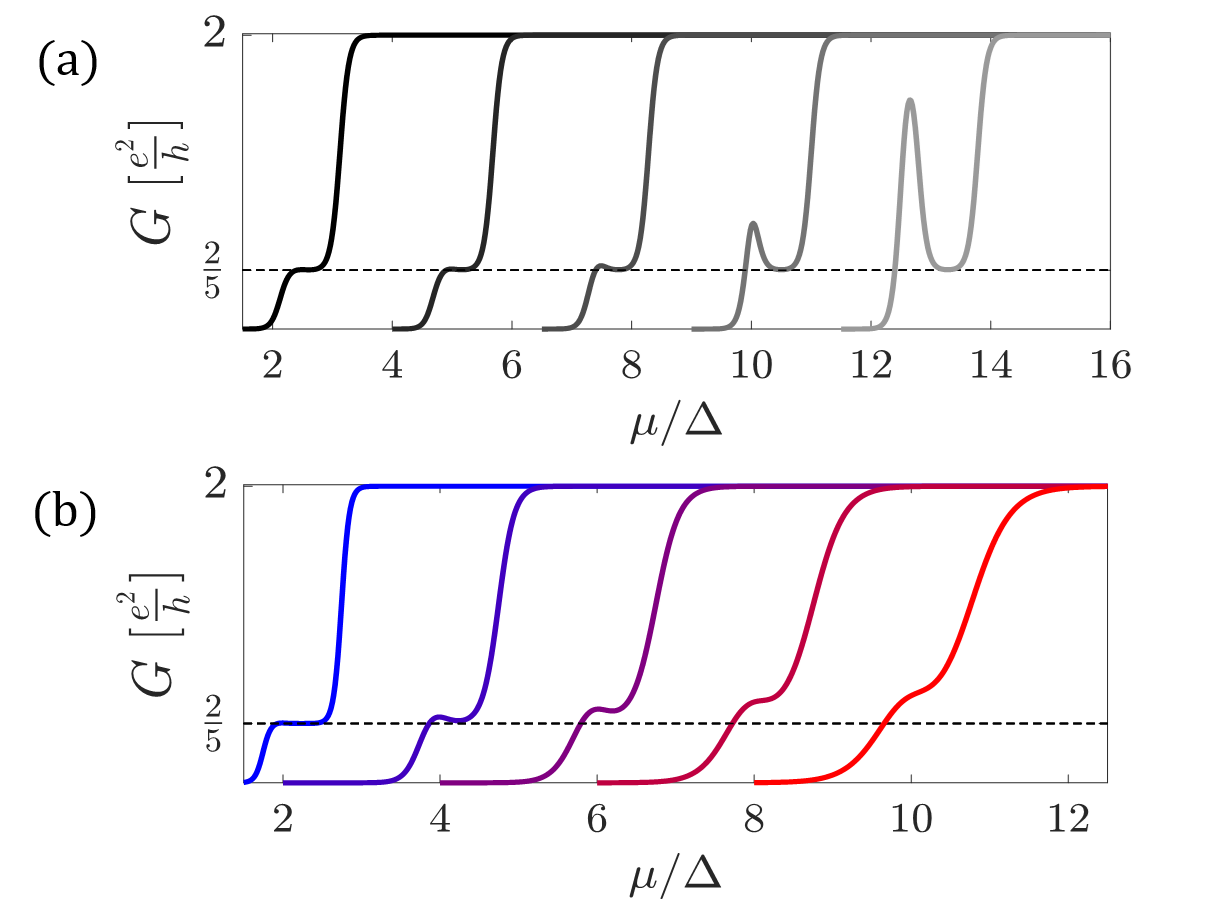}

\caption{\label{fig:FIG2plateaus} Conductance in a gate-voltage sweep for
the time-reversal invariant nano-wire described by the Hamiltonian
density $\mathcal{H}_{R}\left(k\right)=\frac{1}{2m}\left(k+\sigma_{z}\alpha\right)^{2}-\mu$,
around the filling corresponding with $\left(1,3\right)$. (a) Plateaus
at $\frac{T}{\Delta}=0.15$, varying interband separation (left to
right) $\frac{\alpha}{\sqrt{2m\Delta}}=1,1.2,1.5,2,2.5$. (b) With
$\frac{\alpha}{\sqrt{2m\Delta}}=1.4$ and different temperatures (left
to right) $\frac{T}{\Delta}=0.1,0.2,0.3,0.4,0.5$. Plots are shifted
horizontally for clarity.}
\end{figure}

Our result \eqref{eq:jcGeneral} is used to calculate the shape of
conductance plateaus in a typical gate-voltage sweep experiment, by
changing the chemical potential that goes into \eqref{eq:ModelHfermions}.
An example is given in Fig. \ref{fig:FIG2plateaus}, where the shape
of the plateau changes as function of band separation and temperature.
Note that the apparent value of the plateau may \textit{differ from
the universal fractional result \eqref{eq:FracG} at finite temperatures}.

\begin{figure}
\begin{centering}
\includegraphics[scale=0.35]{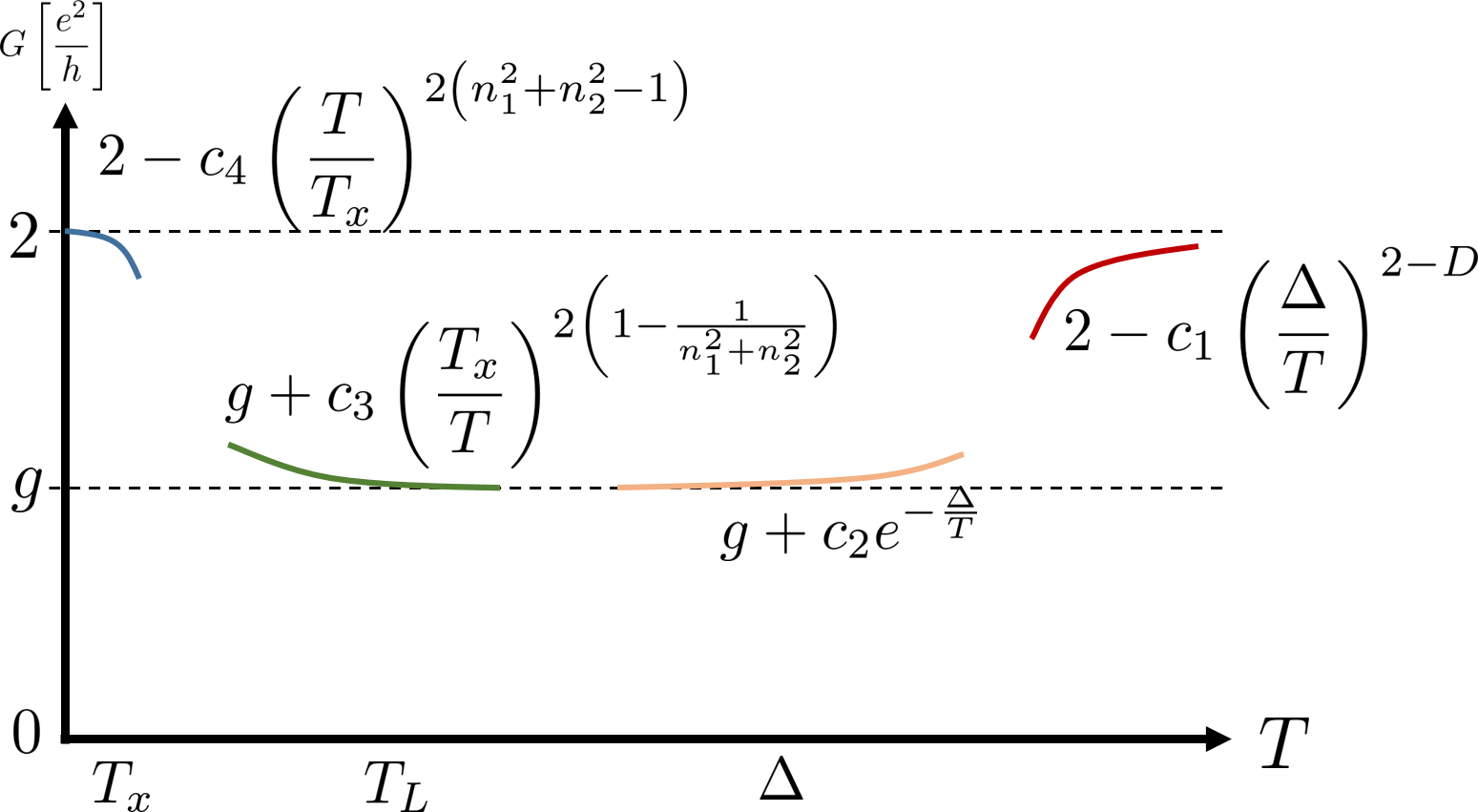}
\par\end{centering}
\caption{\label{fig:FIG3powerlaws-1-1} Schematic depiction of the conductance
as a function of temperature. At temperatures above $T_{L}$ the behavior
is determined by the ratio $\frac{\Delta}{T}$, with an exponentially
small correction to the universal fractional value $g$ at low temperatures,
and a power-law behavior at high temperatures. Below $T_{L}$, the
exponentially small energy scale $T_{x}$ determines the universal
power-law of the conductance.}
\end{figure}

\emph {Ultra-low $T$ limit.---} Our re-fermionization results cease
to be valid for finite system length once the temperature is sufficiently
low, i.e., for $T\ll\frac{u}{L}\equiv T_{L}$, which may be understood
from the following. Upon re-scaling the bosonic fields, one should
in principle also apply the same transformation to the leads, before
matching the boundary conditions. Neglecting this step may by justified,
in the case where all two-point correlators involved in the current,
$\left\langle e^{2i\phi_{g}\left(x,\tau\right)}e^{-2i\phi_{g}\left(x',\tau'\right)}\right\rangle $,
approach their value for a uniform LL. This occurs at $T\gg T_{L}$.
In the opposite limit, we may treat the interacting section as a point-like
perturbation in the non-interacting leads \cite{Ponoma}. Using well-known
results for such perturbations \cite{kane1992transport,KaneImpurity},
we find universal power-law behavior of the conductance \cite{Supplement}.
Slightly below $T_{L}$, the correction to the universally fractionalized
value is

\begin{equation}
G-g\frac{e^{2}}{h}\propto\left(\frac{T_{x}}{T}\right)^{2\left(1-\frac{1}{n_{1}^{2}+n_{2}^{2}}\right)}.\label{eq:TlIncrease}
\end{equation}
Reducing the energy scale further below $T_{x}\sim T_{L}e^{-\Delta/T_{L}}$,
an integer conductance is recovered, behaving much below $T_{x}$
as
\begin{equation}
G-2\frac{e^{2}}{h}\propto\left(\frac{T}{T_{x}}\right)^{2\left(n_{1}^{2}+n_{2}^{2}-1\right)}.\label{eq:Tlto2}
\end{equation}
This result is in agreement with \cite{Ponoma}, who considered the
case $n_{1}=n_{2}=1$. Thus, for higher order backscattering processes,
the conductance will tend to perfect transmission more sharply. The
different $T$-dependent conductance regimes are summarized in Fig.
\ref{fig:FIG3powerlaws-1-1}.

The temperature dependence of the conductance is modified in the presence
of a small amount of sharp impurities. These impurities, which are
more relevant in the RG sense, will impede the flow of $y^{*}$ to
strong coupling and ensure an integer value of the conductance is
not reached, even exactly at $T=0$. The qualitative behavior of the
conductance in the presence of such impurities, and their effects
in the higher temperature limit, are intricate, and depend on the
energy scales $\Delta$, $T_{L}$, and the impurity energy scale,
see SM, Sec. \ref{ImpuritiesSec} \cite{Supplement}.

\emph{Connection with recent experiments.---} The results and discussion
above are particularly interesting, as plateaus which are a fraction
of $\frac{e^{2}}{h}$ have been recently experimentally observed \cite{SelfOrgaanized,FractionsprePublished}.
We conjecture that \textit{weak confinement} in the lateral direction,
a crucial ingredient in obtaining the experimentally observed fractional
plateaus, gives rise to the appearance of additional modes, originating
in transverse direction quantization (cf. a similar argument in \cite{WignerReview}).
Thus, properly tuning a gate, a commensurability condition, which
allows $\mathcal{O}_{\lambda}$ to establish itself, may occur, subsequently
leading to formation of a fractional plateau. Although according to
\eqref{eq:FracG} this would generically lead to $2>g>1$, unlike
the reported measurements, a scenario such as in Fig. \ref{fig:Fig1Schem}d,
where channels may be confined, yet interact strongly throughout the
system, result in \eqref{eq:OtherGs} with $g_{i{\rm c}}<1$, capturing
some of the values obtained experimentally. The \textit{lateral asymmetry}
of the 1D channel, which was found to bear great influence on the
measurements, may also play some role, as it could lead to an effective
SO interaction \cite{AsymmetrySO1,AsymmetrySO2}. If this is indeed
the case, then commensurability may be established between modes of
effective opposing helicity \cite{Supplement}. Notice that contrary
to Rashba nanowires, we find that that plateaus may form \textit{in
the absence of magnetic field}, in a time reversal conserving fashion.
In fact, the most relevant time reversal conserving processes are
with $n_{1}=1$ and $n_{2}=3$ leading to the universal value $g=\frac{2}{5}$,
was observed without magnetic field \cite{PepperArxiv}.

The shape of plateaus that we calculated (Fig. \ref{fig:FIG2plateaus})
fits well to the measurements. Specifically, a conductance peak to
the left of the plateau region is often observed. It is a signature
of the gate-voltage regime lower than the critical commensurate value,
where the conductance should attain its higher, non-fractional value.

Lastly, we comment on the actual values of fractional conductance
that were measured. While some reported values may indeed occur in
our theoretical model, others are absent, e.g., $\frac{2}{3}$, $\frac{3}{10}$.
A plausible explanation is that perhaps some reported plateaus do
not necessarily sit at universal values due to finite temperature,
cf. Fig. \ref{fig:FIG2plateaus}b. Moreover, taking into account impurities,
or having $T\lesssim T_{L}$, the conductance is expected to be non-universal,
albeit maintaining the presence of a chemical potential ``window''
where the fractional conductance value is stabilized, i.e., a plateau.

\emph{Conclusions.---} In this work, we have shown that stabilization
of  fractional two terminal conductance plateaus, which are at a universal
fraction of $\frac{e^{2}}{h}$ depending only on band fillings, requires
sufficiently strong interactions and tuning of the chemical potentials.
In addition to the two-terminal scenario, we also implement our two-band
model to fractional Coulomb drag setups, and to cases where some species
are confined within the wire. Solving the re-fermionized problem exactly
on the generalized Luther-Emery line, we were able to obtain quantitative
finite temperature and length corrections to the universal value,
as well as the restoration of the integer value of the conductance
at ultra small temperatures. This allowed us to suggest a feasible
explanation to recent experimental observation of factional conductance
plateaus. We expect that further insight into these experiments may
be attained from measuring the behavior of the conductance with varying
temperatures, and the shot-noise.

\emph{Acknowledgments.---}  We acknowledge enlightening discussion
with Karsten Flensberg, Sanjeev Kumar, Tommy Li, Yigal Meir, and Michael
Pepper. This work was partially supported by the European Union\textquoteright s
Horizon 2020 research and innovation programme (grant agreement LEGOTOP
No 788715), the DFG (CRC/Transregio 183, EI 519/7- 1), and the Israel
Science Foundation (ISF) and the Binational Science Foundation (BSF).

\bibliographystyle{../auxiliary/apsrev4-1}
\bibliography{../auxiliary/references}

\begin{thebibliography}{31}
\bibitem [{Sup()}]{Supplement}%
  \BibitemOpen
  \href@noop {} {}\bibinfo {note} {Expressions such as \begin {math}
  \left({\psi_r}\right)^m \end {math} are to be understood as \begin {math}
  \psi_r\left(x\right)\psi_r\left(x+a\right)
  ...\psi_r\left(x+\left(m-1\right)a\right) \end {math} due to their fermionic
  nature.}\BibitemShut {Stop}%
\bibitem [{Adi()}]{AdiabaticGap}%
  \BibitemOpen
  \href@noop {} {}\bibinfo {note} {One needs to assume the length scale over
  which the gap is formed is sufficiently larger compared to \begin {math}
  \frac{u}{\Delta}\end{math}, such that fermions with energy higher than the
  gap go through the system with approximately unity transmission.}\BibitemShut
  {Stop}%
\end{thebibliography}

\begin{widetext}

\renewcommand{\thesection}{S.\Alph{section}} \setcounter{figure}{0} \renewcommand{\thefigure}{S\arabic{figure}} \setcounter{equation}{0} \renewcommand{\theequation}{S\arabic{equation}}

\section*{Supplemental Material for ``Fractional Conductance in Strongly Interacting 1D Systems''}

In this supplemental material, we provide technical details for some
of the main results of our work, namely the scattering matrix calculations,
generalized to a many-band scenario, and the re-fermionized conductance
calculations. Additionally, we discuss the consequences of having
time-reversal symmetry in the system, some details of the low-$T$
limit, and how the presence of small disorder impacts our findings.

\section {Scattering matrix calculations and many-bands generalization}\label{NbandsGeneralization}

\subsection {Conductance}

Here we outline the calculation performed in the scattering matrix
formalism for a general case of $N$ species of interacting electrons
in the 1D system. In the main text, the case of $N=2$ was explored.
As mentioned in the main text, we adiabatically attach non-interacting
leads to both ends of the system, and define the incoming vector current
$\vec{I}=\left(I_{R,1},...,I_{R,N},I_{L,1},...,I_{L,N}\right)^{T}$
with $I_{r,i}=\frac{e}{2\pi}\partial_{t}\varphi_{r}^{i}|_{x=r\infty}$,
and similarly the outgoing vector $\vec{O}$. Assuming left-right
symmetry in our system, as well as conservation of current, the incoming
and outgoing currents can be related by
\begin{equation}
\begin{pmatrix}O_{R}\\
O_{L}
\end{pmatrix}=\begin{pmatrix}\mathcal{T} & 1-\mathcal{T}\\
1-\mathcal{T} & \mathcal{T}
\end{pmatrix}\begin{pmatrix}I_{R}\\
I_{L}
\end{pmatrix},
\end{equation}
with $\mathcal{T}$ a $N\times N$ matrix, and we have separated the
current vectors into chiral vectors of length $N$, e.g., $I_{R}=\left(I_{R,1},...,I_{R,N}\right)^{T}$.

Consider the backscattering operator $\left(\psi_{R}^{1\dagger}\psi_{L}^{1}\right)^{n_{1}}\left(\psi_{R}^{2\dagger}\psi_{L}^{2}\right)^{n_{2}}...\left(\psi_{R}^{N\dagger}\psi_{L}^{N}\right)^{n_{N}}$,
with $n_{i}$ integers, and negative $n_{i}$ should be interpreted
as backscattering in the opposite direction, i.e. $\left(\psi_{R}^{1\dagger}\psi_{L}^{1}\right)^{-\left|n\right|}\equiv\left(\psi_{L}^{1\dagger}\psi_{R}^{1}\right)^{\left|n\right|}$.
If for a given $i$ $n_{i}=0$, this species is absent from the backscattering
process (though still possibly contributes to the transport). In the
language of our bosonization scheme, this operator takes the form
\begin{equation}
\lambda_{N}\cos\left(2\sum_{i}n_{i}\phi_{i}\right),\label{eq:ONNN}
\end{equation}
and has a scaling dimension (up to corrections due to inter-species
forward scattering) $\sim\left(\sum_{i}n_{i}^{2}\right)K$, with $K$
the Luttinger parameter accounting for intra-species electron-electron
interactions.

In the limit $\lambda_{N}\rightarrow\infty$ this perturbation pins
$\phi_{g}\equiv\frac{\sum_{i}n_{i}\phi_{i}}{\sqrt{\sum_{i}n_{i}^{2}}}$
to a constant value, leading to the boundary condition
\begin{equation}
\sum_{i}n_{i}\partial_{t}\left(\varphi_{L}^{i}-\varphi_{R}^{i}\right)=0,\label{eq:pinnigN}
\end{equation}
inside the interacting section of the wire. The representation of
$\phi_{g}$ in terms of $n_{i}$ can be thought of as a normalized
vector in $N$-dimensional space, $\phi_{g}=\mathbf{n}\cdot\vec{\phi}$,
with $\vec{\phi}=\left(\phi_{1},...,\phi_{N}\right)^{T}$. Notice
that $\mathbf{n}^{T}\mathbf{n}=1$. Taken at the opposite ends of
the wire, \eqref{eq:pinnigN} gives
\[
\mathbf{n}^{T}\left(O_{L}-I_{R}\right)=0,\,\,\,\,\mathbf{n}^{T}\left(O_{R}-I_{L}\right)=0,
\]
or equivalently,
\[
\mathbf{n}^{T}\mathcal{T}\left(I_{L}-I_{R}\right)=0.
\]
As this result does not depend on the incoming current vector $\vec{I}$,
we find the condition 
\begin{equation}
\mathbf{n}^{T}\mathcal{T}=0.\label{eq:Ncond_pin}
\end{equation}
The remaining gapless modes $\phi_{f,2},...,\phi_{f,N}$ span the
$\left(N-1\right)$-dimensional plane perpendicular to $\phi_{g}$,
such that $\phi_{f,j}\equiv\mathbf{m}_{j}\cdot\vec{\phi}$, and for
all $j$ $\mathbf{m}_{j}\cdot\mathbf{n}=0$. We assume these vectors
are normalized as well, $\mathbf{m}_{j}^{T}\mathbf{m}_{j}=1$ for
all $j$. These modes are assumed to propagate freely throughout the
wire, leading to another $2N-2$ boundary equations, which are written
in terms of the current vectors as 
\[
\mathbf{m}_{j}^{T}\left(O_{R/L}-I_{R/L}\right)=0.
\]
Similarly to before, this yields another condition on the $\mathcal{T}$
matrix,
\begin{equation}
\forall j,\,\,\,\,\,\,\,\,\mathbf{m}_{j}^{T}\left(1-\mathcal{T}\right)=0.\label{eq:Ncond_free}
\end{equation}
It is easily verifiable that 
\begin{equation}
\mathcal{T}=1-\mathbf{n}\mathbf{n}^{T}\label{eq:Tsolution}
\end{equation}
is a solution of \eqref{eq:Ncond_pin},\eqref{eq:Ncond_free}. Since
these boundary conditions fully specify how $\mathcal{T}$ operates
on a complete basis of the $N$-dimensional space (it is spanned by
$\mathbf{n}$ and all the $\mathbf{m}_{j}$ vectors), Eq. \eqref{eq:Tsolution}
is also the \textit{only} solution.

The two-terminal conductance may be extracted by imposing a voltage
difference between the different sides of the system, which amounts
to
\begin{equation}
g=1_{N}^{T}\mathcal{T}1_{N}=N-\frac{\left(\sum_{i}n_{i}\right)^{2}}{\sum_{i}n_{i}^{2}},\label{eq:GgeneralN}
\end{equation}
with $1_{N}$ a column vector of ones of length $N$. This result
reduces to Eq. \eqref{eq:FracG} of the main text in the case of two
fermionic species. Scenarios similar to Fig. \ref{fig:Fig1Schem}d
may also be considered, by attaching only some of the modes to the
voltage leads. The vector $1_{N}$ is replaced by the vector $a_{N}$,
which is comprised of ones for the channels attached to the leads,
and zeros elsewhere, such that Eq. \eqref{eq:GgeneralN} is modified
to
\begin{equation}
g_{{\rm c}}=N_{a}-\frac{\left(\sum_{i\in a}n_{i}\right)^{2}}{\sum_{i}n_{i}^{2}},\label{eq:partialG}
\end{equation}
with $N_{a}$ the number of attached modes and $\sum_{i\in a}$ is
a sum over the coefficients of the attached modes only. As an example,
for $\mathbf{n}=\left(1,-2,1,-2\right)$, if only the second and fourth
modes arrive at the leads, one obtains $g_{{\rm c}}=2-\frac{16}{10}=\frac{2}{5}$.

Some additional examples of fractional conductance coefficients, occurring
for the two band ($N=2$) case are given in Table \ref{tab:Gtable_2band}.

\begin{table}[H]
\begin{centering}
{\large{}}%
\begin{tabular}{ccccccc}
\hline 
\textbf{\large{}$\left(n_{1},\left|n_{2}\right|\right)$} & \textbf{\large{}$g_{n_{1},-\left|n_{2}\right|}$} & \textbf{\large{}$g_{n_{1},\left|n_{2}\right|}$} & $g_{12}$ & \multirow{1}{*}{$g_{i{\rm c}}$} & $\frac{q^{*}}{e}\left(n_{1},-\left|n_{2}\right|\right)$ & $\frac{q^{*}}{e}\left(n_{1},\left|n_{2}\right|\right)$\tabularnewline[\doublerulesep]
\hline 
\hline 
{\large{}$\left(1,1\right)$} & $2$ & $0$ & $\frac{1}{2}$ & $\frac{1}{2}$ & $0$ & $1$\tabularnewline[\doublerulesep]
\hline 
\hline 
{\large{}$\left(1,2\right)$} & $\frac{9}{5}$ & $\frac{1}{5}$ & $\frac{2}{5}$ & $\frac{1}{5}{\rm \,or\,}\frac{4}{5}$ & $\frac{1}{5}$ & $\frac{3}{5}$\tabularnewline[\doublerulesep]
\hline 
\hline 
{\large{}$\left(1,3\right)$} & $\frac{8}{5}$ & $\frac{2}{5}$ & $\frac{3}{10}$ & $\frac{1}{10}{\rm \,or\,}\frac{9}{10}$ & $\frac{1}{5}$ & $\frac{2}{5}$\tabularnewline[\doublerulesep]
\hline 
\hline 
{\large{}$\left(1,4\right)$} & $\frac{25}{17}$ & $\frac{9}{17}$ & $\frac{4}{17}$ & $\frac{1}{17}{\rm \,or\,}\frac{16}{17}$ & $\frac{3}{17}$ & $\frac{5}{17}$\tabularnewline[\doublerulesep]
\hline 
\hline 
{\large{}$\left(2,3\right)$} & $\frac{25}{13}$ & $\frac{1}{13}$ & $\frac{6}{13}$ & $\frac{4}{13}{\rm \,or\,}\frac{9}{13}$ & $\frac{1}{13}$ & $\frac{5}{13}$\tabularnewline[\doublerulesep]
\hline 
\end{tabular}{\large\par}
\par\end{centering}
\caption{\label{tab:Gtable_2band} Examples of the different fractional transport
coefficients. The second and third columns correspond to total momentum
conserving ($\mathcal{O}_{{\rm bs}}$) or umklapp-like ($\mathcal{O}_{{\rm um}}$)
processes. The fourth column is the drag transconductance. The conductance
$g_{i{\rm c}}$ is obtained when $n_{1}$ or $n_{2}$ bands do not
reach the voltage leads. The last two columns are the corresponding
Fano factors obtained from the tunneling shot-noise \eqref{eq:FanoFactor}.}
\end{table}

\subsection {Tunneling shot-noise}

For a system with a finite length, tunneling events of charge between
the non-interacting leads may affect the conductance. Such an event
is represented by tunneling between adjacent minima of the cosine
of \eqref{eq:ONNN}. These minima are fixed in the limit $L\rightarrow\infty$,
$T=0$. Tunneling between adjacent minima causes a $\frac{\pi}{\sqrt{\sum_{i}n_{i}^{2}}}$
temporal kink in $\phi_{g}$ for the duration of the tunneling. The
total charge transferred between the leads can be easily found by
integrating the charge current over the time of the tunneling event.
Since the charge current is given by
\[
j_{c}=\frac{1}{\pi}\partial_{t}\sum_{i}\phi_{i}=\frac{1}{\pi}\partial_{t}\left[\left(\mathbf{n}\cdot1_{N}\right)\phi_{g}+\sum_{j}\left(\mathbf{m}_{j}\cdot1_{N}\right)\phi_{f,j}\right],
\]
 the total fractional charge transferred is

\begin{equation}
q^{*}=\int dtj_{c}=\frac{\sum_{i}n_{i}}{\sum_{i}n_{i}^{2}}e.\label{eq:FanoFactor}
\end{equation}
For $V\gg T,T_{L}$, these tunneling events will dominate the dc shot-noise
given by \cite{ChamonNoise} 
\begin{equation}
S\left(\omega\rightarrow0\right)=2q^{*}I_{t},
\end{equation}
with $I_{t}$ the excess tunneling current in the $\phi_{g}$ channel,
$I_{t}=I-\frac{e^{2}}{h}gV$, with $I$ being the total measured current.
Thus, we identify $q^{*}$ with the Fano factor of this shot-noise
contribution. This result is a many-band generalization of a similar
formula obtained in previous works \cite{SelaShotNoise}, and gives
the same result for the $N=2$ case.

\section {Detailed derivation of the refermionization solution} \label{RefermionizationDetails}

Let us write the Euclidean action accounting for \eqref{eq:LuttingerPM}--\eqref{eq:GenOnm}
as
\begin{equation}
S=\frac{u}{2\pi}\int dxd\tau\left[\mathcal{L}_{f}+\mathcal{L}_{g}+\mathcal{L}_{\mathcal{O}}+\mathcal{L}_{\times}\right],\label{eq:EuclidAction}
\end{equation}
with 
\[
\mathcal{L}_{a}=\phi_{a}\left(-\frac{1}{K_{a}^{2}}\partial_{x}^{2}-\frac{1}{u^{2}}\partial_{\tau}^{2}\right)\phi_{a},
\]
for $a=f,g$, 
\[
\mathcal{L}_{\mathcal{O}}=y\cos\left(2\sqrt{n_{1}^{2}+n_{2}^{2}}\phi_{g}\right),
\]
\[
\mathcal{L}_{\times}=\frac{1}{K_{\times}^{2}}\partial_{x}\phi_{f}\partial_{x}\phi_{g},
\]
and the modified Luttinger parameters
\[
K_{f}^{-2}=\frac{1}{2}\left(\frac{1}{K_{+}^{2}}\frac{\left(n_{1}-n_{2}\right)^{2}}{n_{1}^{2}+n_{2}^{2}}+\frac{1}{K_{-}^{2}}\frac{\left(n_{1}+n_{2}\right)^{2}}{n_{1}^{2}+n_{2}^{2}}\right),
\]
 
\[
K_{g}^{-2}=\frac{1}{2}\left(\frac{1}{K_{+}^{2}}\frac{\left(n_{1}+n_{2}\right)^{2}}{n_{1}^{2}+n_{2}^{2}}+\frac{1}{K_{-}^{2}}\frac{\left(n_{1}-n_{2}\right)^{2}}{n_{1}^{2}+n_{2}^{2}}\right),
\]
\[
K_{\times}^{-2}=\frac{n_{1}^{2}-n_{2}^{2}}{n_{1}^{2}+n_{2}^{2}}\left(\frac{1}{K_{-}^{2}}-\frac{1}{K_{+}^{2}}\right).
\]
The cross term $\mathcal{L}_{\times}$ should not be neglected here
(the way it implicitly was in the scattering matrix calculations),
as the boundary conditions for the voltage leads \eqref{eq:BCradiative}
contain $K_{\pm}$, which account for the electrostatic charging of
the interacting wire \cite{LeadsBC}. Ignoring $\mathcal{L}_{\times}$would
thus be inconsistent with \eqref{eq:BCradiative} and lead to non-universal
asymptotic conductance. By a shift of $\phi_{f}\rightarrow\phi_{f}+Q\phi_{g}$,
with $Q$ chosen as
\begin{equation}
Q\equiv\frac{\left(n_{1}^{2}-n_{2}^{2}\right)\left(\frac{1}{K_{+}^{2}}-\frac{1}{K_{-}^{2}}\right)}{\frac{1}{K_{-}^{2}}\left(n_{1}-n_{2}\right)^{2}+\frac{1}{K_{-}^{2}}\left(n_{1}+n_{2}\right)^{2}},
\end{equation}
we find the modified Lagrangian densities
\begin{align}
\mathcal{L}_{g}+\mathcal{L}_{\mathcal{O}} & =\phi_{g}\left(-\left(K_{g}^{-2}+K_{f}^{-2}Q^{2}\right)\partial_{x}^{2}-\frac{1+Q^{2}}{u^{2}}\partial_{\tau}^{2}\right)\phi_{g}\nonumber \\
 & +y\cos\left(2\sqrt{n_{1}^{2}+n_{2}^{2}}\phi_{g}\right),
\end{align}
\begin{align}
\mathcal{L}_{\times} & =\frac{2Q}{u^{2}}\partial_{\tau}\phi_{fLL}\partial_{\tau}\phi_{g}.
\end{align}
Notice that $\mathcal{L}_{f}$ is unaffected by this transformation.
As $\mathcal{L}_{\times}$ now vanishes in the static limit, it will
henceforth be neglected in the massive $\phi_{g}$ regime. With the
shift performed above, we write the current and densities operators
using 
\[
\hat{O}_{+}=\frac{\left[\left(n_{1}+n_{2}\right)+Q\left(n_{2}-n_{1}\right)\right]\hat{O}_{g}+\left(m-n\right)\hat{O}_{f}}{\sqrt{n_{1}^{2}+n_{2}^{2}}},
\]
\[
\hat{O}_{-}=\frac{\left[\left(n_{1}-n_{2}\right)+Q\left(n_{2}+n_{1}\right)\right]\hat{O}_{g}+\left(m+n\right)\hat{O}_{f}}{\sqrt{n_{1}^{2}+n_{2}^{2}}},
\]
with $\hat{O}=\rho,j$. These expressions are plugged in \eqref{eq:BCradiative}
to obtain the boundary equations in terms of the $g$ and $f$ bosonic
fields.

Before the re-fermionization step, we rescale the bosonic fields,
\[
\tilde{\phi}_{f}=\frac{1}{\sqrt{K_{f}}}\phi_{f},\,\,\,\,\,\tilde{\theta}_{f}=\sqrt{K_{f}}\theta_{f},
\]
\[
\tilde{\phi}_{g}=\frac{1}{\sqrt{\tilde{K}}}\phi_{g},\,\,\,\,\,\tilde{\theta}_{g}=\sqrt{\tilde{K}}\theta_{g},
\]
with $\tilde{K}=\sqrt{\frac{1+Q^{2}}{K_{g}^{-2}+K_{f}^{-2}Q^{2}}}$.
At the special line defined by $\tilde{K}=\frac{1}{n_{1}^{2}+n_{2}^{2}}\equiv K^{*}$,
the gapped channel describes non-interacting fermions (the $\tilde{f}$
sector is free as well, for any $K_{f}$). The quadratic in fermion
operators Hamiltonian may be written as

\begin{align}
H & =\int dx\Psi_{f}^{\dagger}\left[iu\sigma_{z}\partial_{x}\right]\Psi_{f}\nonumber \\
 & +\int dx\Psi_{g}^{\dagger}\left[iu\sigma_{z}\partial_{x}+\Delta\left(x\right)\sigma_{x}\right]\Psi_{g},\label{eq:DiracLE}
\end{align}
with$\Psi_{g}=\left(L_{g},R_{g}\right)^{T},\,\,\,\,\,\Psi_{f}=\left(L_{f},R_{f}\right)^{T},$
and the chiral fermionic fields defined as vertex operators of the
rescaled bosonic variables, $R_{j}\sim e^{-i\left(\tilde{\phi}_{j}-\tilde{\theta}_{j}\right)},$
$L_{j}\sim e^{i\left(\tilde{\phi}_{j}+\tilde{\theta}_{j}\right)}$.
The density and current operators of the two sectors are thus given
by 
\begin{equation}
\rho_{j}=\Psi_{j}^{\dagger}\Psi_{j},\label{eq:rhodef}
\end{equation}
\begin{equation}
j_{j}=-u\Psi_{j}^{\dagger}\sigma_{z}\Psi_{j},\label{eq:jdef}
\end{equation}
and we may express the boundary conditions in terms of them. Next,
we look for solutions for the Schrodinger equation $H\Psi_{j}=E\Psi_{j}$.
We find 
\begin{equation}
\Psi_{f}\left(E\right)=\begin{pmatrix}\eta_{1}e^{i\frac{E}{u}\left(x-\frac{L}{2}\right)}\\
\eta_{2}e^{-i\frac{E}{u}\left(x+\frac{L}{2}\right)}
\end{pmatrix},\label{eq:PsibSol-1}
\end{equation}
with $\eta_{1/2}$ fermionic operators. Clearly, by using \eqref{eq:rhodef}--\eqref{eq:jdef},
such a solution gives rise to spatially independent forms of $j_{f},\rho_{f}$.
Assuming the spatial profile of the gap $\Delta\left(x\right)$ is
sufficiently smooth at the connection to the leads, i.e., varies on
a length scale greater than $\frac{u}{\Delta}$, we may assume a similar
form for the gapped fermions wave function above the gap (as backscattering
is suppressed),
\begin{equation}
\Psi_{g}\left(E>\Delta\right)=\begin{pmatrix}\eta_{3}e^{i\frac{\sqrt{E^{2}-\Delta^{2}}}{u}\left(x-\frac{L}{2}\right)}\\
\eta_{4}e^{-i\frac{\sqrt{E^{2}-\Delta^{2}}}{u}\left(x+\frac{L}{2}\right)}
\end{pmatrix},\label{eq:Psigsol_balist-1}
\end{equation}
which again results in spatially uniform charge and current densities.
Thus, for $E>\Delta$, we may solve \eqref{eq:BCradiative} as a set
of equations for four position independent variables, and extract
\begin{equation}
j_{c}\left(E>\Delta\right)=\frac{1}{\pi}\delta f.\label{eq:jc1}
\end{equation}

For energies below the gap, we find an exponentially decaying solution
along the system, 
\begin{align}
\Psi_{g}\left(E<\Delta\right) & =\frac{1}{\sqrt{2}}\xi_{1}\begin{pmatrix}1\\
\frac{E-i\kappa u}{\tilde{\Delta}}
\end{pmatrix}e^{-\kappa\left(x+\frac{L}{2}\right)}\nonumber \\
 & +\frac{1}{\sqrt{2}}\xi_{2}\begin{pmatrix}1\\
\frac{E+i\kappa u}{\tilde{\Delta}}
\end{pmatrix}e^{\kappa\left(x-\frac{L}{2}\right)},\label{eq:PsigSolGapped-1}
\end{align}
with $\kappa\equiv\frac{\sqrt{\Delta^{2}-E^{2}}}{u}$, and fermionic
operators $\xi_{1/2}$. Thus, we may express the charge and current
operators of the gapped re-fermions as
\begin{align}
j_{g} & =-ue^{-\kappa L}\left(\frac{\kappa u}{\kappa u+iE}\xi_{1}^{\dagger}\xi_{2}+\mathrm{h.c.}\right),\label{eq:jManipulate-1}
\end{align}
\begin{align}
\rho_{g} & \left(x=-\frac{L}{2}\right)=\xi_{1}^{\dagger}\xi_{1}+\xi_{2}^{\dagger}\xi_{2}e^{-2\kappa L}+e^{-\kappa L}\left(\frac{\xi_{1}^{\dagger}\xi_{2}}{1-i\frac{\kappa u}{E}}+\mathrm{h.c.}\right),
\end{align}
\begin{align}
\rho_{g} & \left(x=+\frac{L}{2}\right)=\xi_{1}^{\dagger}\xi_{1}e^{-2\kappa L}+\xi_{2}^{\dagger}\xi_{2}+e^{-\kappa L}\left(\frac{\xi_{1}^{\dagger}\xi_{2}}{1-i\frac{\kappa u}{E}}+\mathrm{h.c.}\right).
\end{align}
Assuming the non-interacting leads are adiabatically connected to
the wire yields an additional boundary condition, $\left\langle R_{g}^{\dagger}\left(-\frac{L}{2}\right)L_{g}\left(\frac{L}{2}\right)\right\rangle =\left\langle \left\langle L_{g}^{\dagger}\left(\frac{L}{2}\right)R_{g}\left(-\frac{L}{2}\right)\right\rangle \right\rangle =0$,
such that there is no $2k_{F}$ backscattering in the leads. This
amounts to the following relations between the fermionic operators,
\begin{equation}
\xi_{1}^{\dagger}\xi_{1}+\xi_{2}^{\dagger}\xi_{2}=-\frac{\kappa u}{E}\cosh\kappa L\left(i\xi_{1}^{\dagger}\xi_{2}\frac{\kappa u-iE}{\kappa u}+\mathrm{h.c.}\right),
\end{equation}
\begin{equation}
\xi_{2}^{\dagger}\xi_{2}-\xi_{1}^{\dagger}\xi_{1}=\sinh\kappa L\left(\xi_{1}^{\dagger}\xi_{2}\frac{\kappa u-iE}{\kappa u}+\mathrm{h.c.}\right).\label{eq:jManipulate2-1}
\end{equation}
Manipulating Eqs. \eqref{eq:jManipulate-1}--\eqref{eq:jManipulate2-1},
we may finally relate the current $j_{g}$ to the \textit{difference
in densities between the ends of the wire},

\begin{align}
j_{g} & =u\frac{\Delta^{2}-E^{2}}{2\Delta^{2}\sinh^{2}\kappa L}\left(\rho_{g}\left(x=+\frac{L}{2}\right)-\rho_{g}\left(x=-\frac{L}{2}\right)\right).\label{eq:mainTunnel}
\end{align}
We may now solve once again \eqref{eq:BCradiative} as a set of linear
equations, but now the variables are $\rho_{f},j_{f},\rho_{g}\left(x=\pm\frac{L}{2}\right)$.
Straightforward calculation yields $j_{f}$ and $\rho_{g}$, and thus
$j_{g}$. We plug them into the total charge current, given in the
``shifted'' basis by 
\begin{equation}
j_{c}=\frac{\left[\left(n_{1}+n_{2}\right)+Q\left(n_{2}-n_{1}\right)\right]j_{g}+\left(n_{2}-n_{1}\right)j_{f}}{\sqrt{n_{1}^{2}+n_{2}^{2}}},
\end{equation}
and we finally obtain after some elaborate yet straightforward manipulations,
\begin{equation}
j_{c}\left(E<\Delta\right)=\frac{1}{2\pi}\delta f\frac{\left(n_{2}-n_{1}\right)^{2}+2\chi}{n_{1}^{2}+n_{2}^{2}+\chi},\label{eq:jc2}
\end{equation}
with the non-universal factor $\chi=\frac{\Delta^{2}-E^{2}}{2\Delta^{2}\sinh^{2}\kappa L}\left(K_{-}^{2}\left(n_{1}-n_{2}\right)^{2}+K_{+}^{2}\left(n_{1}+n_{2}\right)^{2}\right)$.
For low enough temperatures, and in the limit $\kappa L\rightarrow0$,
one finds $\chi\rightarrow\infty$, and an integer conductance of
$g=2$ is restored. For the opposite limit, $\kappa L\rightarrow\infty$,
one recovers the universal value, Eq. \eqref{eq:FracG}. The dependence
of the conductance on temperature, chemical potential (i.e., the distance
from the commensurability point), and voltage, are encapsulated within
the $\delta f$ dependence.

Combining Eqs. \eqref{eq:jc1},\eqref{eq:jc2}, integrating over energy,
and restoring units, we may calculate the two-terminal conductance
for arbitrary temperature and system length. An example is shown in
Fig. \ref{fig:suppColormapcuts} for the case of $\left(1,3\right)$,
which was considered in Fig. \ref{fig:FIG2plateaus}. Additionally,
we show two cuts with constant temperature or length, showing the
power-law behavior at small $L$ and high $T$.

\begin{figure}
\begin{centering}
\includegraphics[scale=0.5]{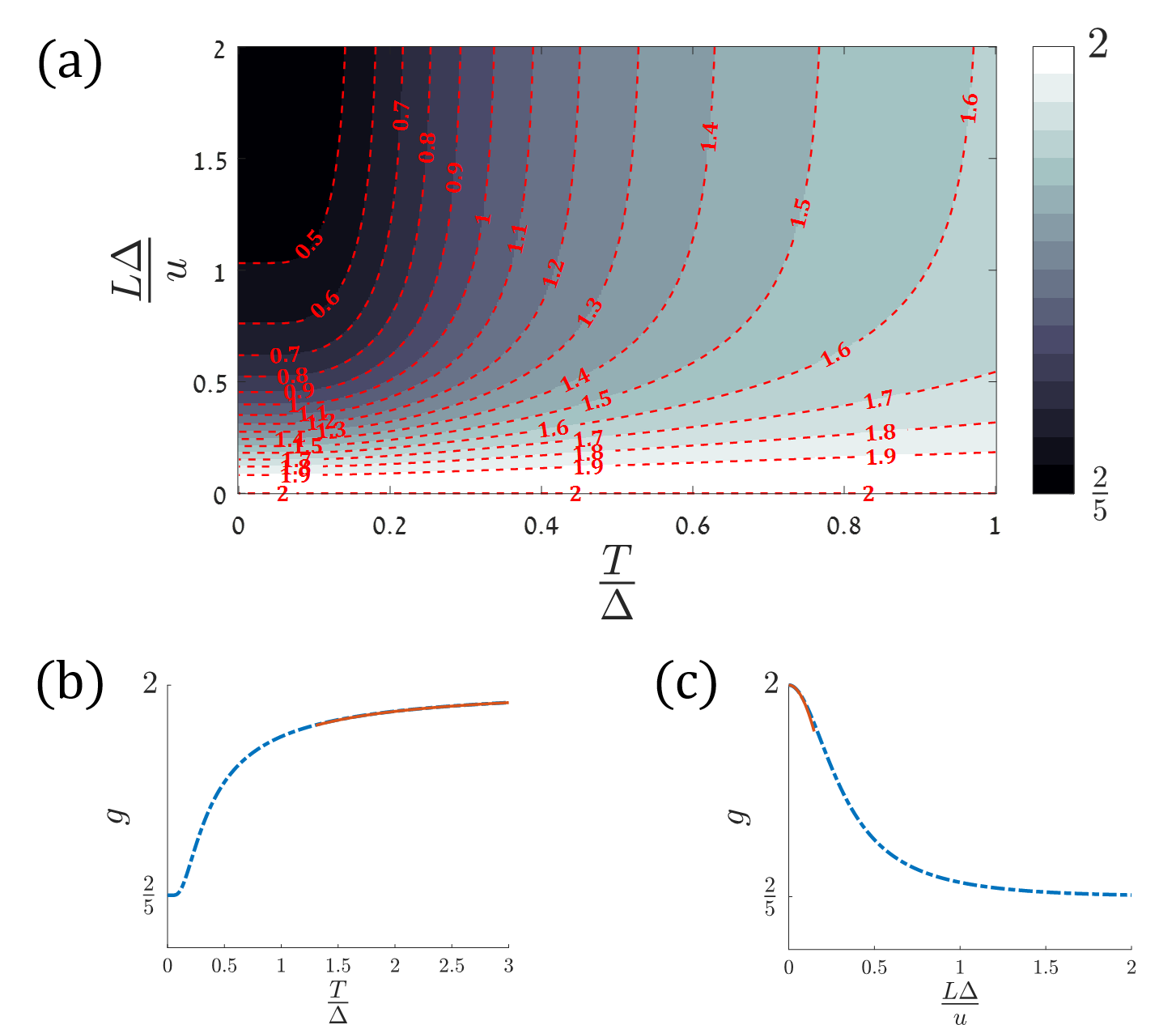}
\par\end{centering}
\caption{\label{fig:suppColormapcuts} Calculation of the conductance (in units
of $\frac{e^{2}}{h}$) for the case of $\left(n_{1},n_{2}\right)=\left(1,3\right)$.
(a) Conductance as a function of $T$ and $L$. (b) Dashed blue line:
a cut with constant $\frac{L\Delta}{u}=1.3$; solid red line: power-law
fit with $2-g\propto\left(\Delta/T\right)$. (c) Dashed blue line:
a cut with constant $\frac{T}{\Delta}=0.15$; solid red line: power-law
fit with $2-g\propto\left(L\Delta/u\right)^{2}$. All calculations
were made with the chemical potential exactly at the $\left(1,3\right)$
commensurability point.}
\end{figure}

\section {Time-reversal invariant systems}\label{TrSec}

Let us consider a system comprised of two one-dimensional channels
of opposite helicities, strongly interacting with one another. The
helicity need not necessarily correspond to the spin itself, but to
a general pseudo-spin degree of freedom, which will be denoted as
$\uparrow/\downarrow$ for convenience. We number each helical channel
by $1,2$ , as in the main text, corresponding to the mapping of the
chiral fermionic operators
\[
\psi_{R}^{1}\leftrightarrow\psi_{R,\uparrow},\,\,\,\,\,\psi_{L}^{1}\leftrightarrow\psi_{L,\downarrow},
\]
\[
\psi_{R}^{2}\leftrightarrow\psi_{R,\downarrow},\,\,\,\,\,\psi_{L}^{2}\leftrightarrow\psi_{L,\uparrow}.
\]
By applying different chemical potentials to the two helical channels,
the results we obtained in the main text may be applied to such a
system.

The presence of time-reversal symmetry modifies the allowed integers
that go into the operator $\mathcal{O}_{\lambda}$. To see this, consider
that under time-reversal the chiral fermionic operators transform
as
\begin{equation}
\psi_{R,\downarrow}\rightarrow\psi_{L,\downarrow},\,\,\,\,\,\psi_{L,\downarrow}\rightarrow-\psi_{R,\downarrow},\,\,\,\,\psi_{L,\uparrow}\rightarrow\psi_{R,\downarrow},\,\,\,\,\,\psi_{R,\downarrow}\rightarrow-\psi_{L,\uparrow}.\label{eq:TRtransform}
\end{equation}
Thus, one finds that $\mathcal{O}_{\lambda}$ is time-reversal invariant
only if $\left(n_{1}+n_{2}\right)$ is an \textit{even }integer.

This scenario may be realized in two different ways, depicted in Fig.
\ref{fig:suppTRschemes}. (i) Using a narrow sample of a two-dimensional
topological insulator (TI), with width $d$ much greater than the
characteristic correlation length $\xi$, with different gate voltages
applied to the different edges, as to achieve the fractional commensurability
of the Fermi momenta. (ii) Constructing a TI-Insulator-TI heterostructure,
with different top and bottom gates, or different doping for the two
topologically non-trivial layers. In the two scenarios one must ensure
that the distance between the different edge states is such that strong
electron-electron interactions may take place. Alternatively, the
physics of a Rashba nano-wire may be considered.

\begin{figure}
\begin{centering}
\includegraphics[scale=0.45]{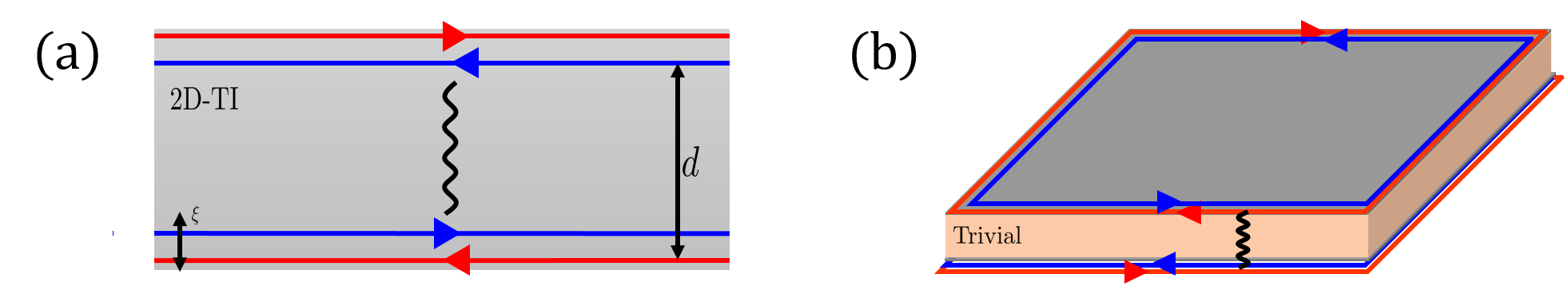}
\par\end{centering}
\caption{\label{fig:suppTRschemes} More robust TR symmetric setups using 2D-TIs.
(a) Edge states of a thin 2D-TI, with its edges kept in different
chemical potential. (b) Edge states of two 2D-TIs with different helicities
separated by a trivial insulator are governed by the same Hamiltonian.}
\end{figure}

\subsection {Rashba nanowire}

The model for a spinfull 1D system with Rashba type spin-orbit coupling
(RSO) is captured by the Hamiltonian density 
\begin{equation}
\mathcal{H}_{R}\left(k\right)=\frac{k^{2}}{2m}+\mu+\alpha\sigma^{z}k+E_{{\rm Z}}\sigma^{y},\label{eq:RashbaDensity}
\end{equation}
with $k$ the wave vector, $\alpha$ the RSO strength, $E_{{\rm Z}}$
the Zeeman energy, and $\sigma^{i}$ the Pauli matrices acting on
the electrons spin degree of freedom. At $E_{{\rm Z}}=0$, Eq. \eqref{eq:RashbaDensity}
describes two copies of parabolic dispersion corresponding to the
value of $\sigma_{z}$, shifted in momentum space by the RSO. Focusing
on the regime below the energy at which the two bands cross, we linearize
the spectrum to obtain the chiral fermion modes, resulting in a system
with two channels of opposite helicity, as discussed above.

Previous studies of fractional helical wires \cite{HelicalSelaOreg,LossHelical}
discussed processes analogous to $\left(n,n+1\right)$ backscattering,
which inherently break time-reversal symmetry and are only generated
in the presence of a finite magnetic field with finite Zeeman energy
$E_{{\rm Z}}$. Our treatment generalizes those results to a variety
of commensurate filling factors, given by $\nu\equiv\frac{k_{F}}{\alpha m}=\frac{n_{1}-n_{2}}{n_{1}+n_{2}}$.
We find that the lowest order non-trivial time-reversal invariant
fractional phase occurs at $\left(1,3\right)$, or $\nu=\frac{1}{2}$,
with a novel fractional conductance $g=\frac{2}{5}$. A fractional
conductance value of $\frac{1}{5}$, found to be the most relevant
in the time-reversal breaking model, may still be obtained for the
filling factor $\nu=\frac{1}{3}$, but it requires a higher order
$\left(2,4\right)$ process for its gap to be established in the system,
and thus stronger interactions.

\section {Ultra-low $T$ limit}\label{lowTsupp}

Our re-fermionization results cease to be valid for finite system
length once the temperature is sufficiently low, i.e., for $T\ll\frac{u}{L}\equiv T_{L}$.
This may be understood from the following. Upon rescaling the bosonic
fields, one should in principle also apply the same transformation
to the voltage leads, before matching the boundary conditions. Neglecting
this step may by justified, in the case where all two-point correlators
involved in the current, $\left\langle e^{2i\phi_{g}\left(x,\tau\right)}e^{-2i\phi_{g}\left(x',\tau'\right)}\right\rangle $,
approach their value for a uniform LL. This occurs at $T\gg T_{L}$.
In the opposite limit, we have to treat the interacting section as
a point-like perturbation in the non-interacting Fermi liquid which
comprises the leads \cite{Ponoma}. The corresponding Hamiltonian
for $\phi_{g}$ in our regime of interest, $\Delta>T,T_{L}$, is given
by 
\begin{equation}
H_{L}=\frac{u}{2\pi}\int dx\left[\left(\partial_{x}\phi_{g}\right)^{2}+\left(\partial_{x}\theta_{g}\right)^{2}+y^{*}\cos\left(\frac{2\theta_{g}}{\sqrt{n_{1}^{2}+n_{2}^{2}}}\right)\delta\left(x\right)\right],\label{eq:LowTsHORT-1-1}
\end{equation}
with the new parameter $y^{*}\approx\frac{T_{L}}{u\sinh^{2}\left(\frac{\Delta}{u}L\right)}$
\cite{Ponoma}. Notice that $y^{*}$ is exponentially small in $\frac{\Delta}{T_{L}}$.
Eq. \eqref{eq:LowTsHORT-1-1} is written in the strong interaction
limit, where $y^{*}$ represents a tunneling event between two semi-infinite
Luttinger liquids. The perturbation $y^{*}$ is clearly relevant in
an RG sense, ensuring it reaches the strong coupling limit at low
enough temperatures $T<t\left(\frac{uy^{*}}{t}\right)^{\frac{n_{1}^{2}+n_{2}^{2}}{n_{1}^{2}+n_{2}^{2}-1}}\equiv T_{x}$.
At $T=0$ this sector becomes perfectly transmitting, and a total
conductance of $2\frac{e^{2}}{h}$ is restored. The Hamiltonian \eqref{eq:LowTsHORT-1-1}
allows us to find power-law behavior in the deviations from the universal
fractional conductance value \eqref{eq:FracG} in the regime $T_{x}\ll T<T_{L}$.
Mapping the problem into that of a strong impurity in an interacting
LL would reveal the perturbative (in $y^{*}$) result \cite{kane1992transport,KaneImpurity}
\begin{equation}
G-g\frac{e^{2}}{h}\propto\left(\frac{T_{x}}{T}\right)^{2\left(1-\frac{1}{n_{1}^{2}+n_{2}^{2}}\right)}.\label{eq:TlIncrease-1}
\end{equation}
By examining the dual model of \eqref{eq:LowTsHORT-1-1}, which has
the dual perturbatively small sine-Gordon term containing $\phi_{g}$,
the power-law deviation from perfect transmission around $T=0$ is
similarly recovered,
\begin{equation}
G-2\frac{e^{2}}{h}\propto\left(\frac{T}{T_{x}}\right)^{2\left(n_{1}^{2}+n_{2}^{2}-1\right)}.\label{eq:Tlto2-1}
\end{equation}

\section {Effect of impurities}\label{ImpuritiesSec}

The 1D system we describe in this work is generally not protected
from the presence of disorder and impurity scattering. We thus explore
under what conditions do such elements spoil the fractional two terminal
conductance and to what extent. In the regime where the scattering
mean-free-path is comparable to the system size it is sufficient to
consider the effect of a single impurity scattering center.

Backscattering of a single particle in the $i$ fermionic channel
is described by an operator $\mathcal{B}_{i}\sim y_{{\rm imp}}\cos2\phi_{i}$.
Its scaling dimension, $\frac{1}{2}\left(K_{+}+K_{-}\right)$, will
generically be smaller than one (making it relevant in the RG sense)
when $\mathcal{O}_{\lambda}$ is relevant, hence the lack of protection
mentioned. However, since the impurity is localized in space, whereas
$\mathcal{O}_{\lambda}$ operates along the entire system, the latter
may grow much faster under the RG flow, and reach strong coupling
first. This is our regime of interest, since it will lead to clear
signatures of the partially gapped state. By very crudely estimating
$K_{+}\approx K_{-}\equiv K$, this happens for $K<\left(n_{1}^{2}+n_{2}^{2}-1\right)^{-1}$,
i.e., repulsive interactions substantially stronger compared to the
interaction required to achieve the situation when the dimension of
the operator ${\cal O}_{\lambda}$, $D<2$, which is equivalent to
$K<2\left(n_{1}^{2}+n_{2}^{2}\right)^{-1}$. Notice that once $\lambda\rightarrow\infty$,
$\phi_{g}$ ``freezes out'' , causing $\mathcal{B}_{i}$ to become
even more relevant as its dimension effectively becomes $D_{{\rm imp}}=\left(1-\frac{n_{i}^{2}}{n_{1}^{2}+n_{2}^{2}}\right)K_{f}$.

We now have two different temperature scales in our problem: $T^{*}=\Delta$,
the gap originating in $\mathcal{O}_{\lambda}$ , and $T_{f}$, associated
with the RG flow of $\mathcal{B}_{i}$, with $T_{f}<T^{*}$ assumed.
For $T\gg T^{*}$, the impurity has an insignificant effect and the
power-law correction to $2e^{2}/h$ are as in \eqref{eq:deltaGpowerLaw}.
At the vicinity of $T^{*}$ and below it, the conductance settles
at the fractional result \eqref{eq:FracG}, with exponentially small
corrections. As the temperature is lowered even further, the impurity
scattering begins to hinder the conductance, until completely gapping
out $\phi_{f}$ as well as $\phi_{g}$ at $T\ll T_{f}$. At these
very low temperatures, one must start considering the additional energy
scale $T_{L}$, and the picture becomes much more complicated. We
will henceforth assume for simplicity that the impurity acts simultaneously
on both the original channels, $\phi_{1},\phi_{2}$.

If $T_{f}<T_{L}<T^{*}$, the impurity never reaches the strong coupling
regime. The impurity contributes a small power-law correction to the
conductance, which behaves as $\left(\frac{T_{f}}{T}\right)^{2\left(1-D_{{\rm imp}}\right)}$
above $T_{L}$, and remains a temperature-independent constant below
$T_{L}$.

On the other hand, in the regime $T_{L}<T_{f}<T_{*}$, the conductance
for temperatures below $T_{f}$ yet significantly above $T_{L}$ will
vanish with a non-universal power-law, as $\left(\frac{T}{T_{f}}\right)^{2\left(1-\frac{1}{D_{{\rm imp}}}\right)}$.
Once again, below $T_{L}$ the small conductance due to the strong
impurity, will remain constant.

Lastly, we note that $T_{f}$ may be ``pushed down'' to lower temperatures,
such that the intermediate temperature regime with conductance very
closed to its fractional universal value is greatly expanded, if the
1D system consists of time-reversal (TR) symmetry protected edge states
of a 2D topological insulator (e.g., tungsten ditelluride \cite{WTE2monolayer,QSH100K})
of opposite helicities. Prohibiting single particle backscattering,
operators of the order $\sim y_{{\rm imp}}\cos4\phi_{i}$ or higher
may be relevant. The scenario in which $\phi_{g}$ is gapped out before
the impurity reaches its strong coupling regime is now roughly given
by $K<\left(n_{1}^{2}+n_{2}^{2}-4\right)^{-1}$, i.e., we require
much weaker interaction strengths for our regime of interest. The
value of $T_{f}\propto y_{{\rm imp}}^{\frac{1}{1-D_{{\rm imp}}}}$
is significantly reduced in this case, since $D_{{\rm imp}}=4\left(1-\frac{n_{i}^{2}}{n_{1}^{2}+n_{2}^{2}}\right)K_{f}$
is four times larger compared to the non-time-reversal-protected system,
and the vanishing conductance power-laws are modified accordingly.

\end {widetext}
\end{document}